\documentclass{PoS}

\title{Matter Bounce Scenario in F(T) gravity}

\ShortTitle{Matter Bounce Scenario in F(T) gravity}

\author{\speaker{Jaume Haro} \thanks{Investigation 
supported by MINECO, project MTM2011-27739-C04-01 and 
MTM2012-38122-C03-01 .}\\
        Universitat Polit\`ecnica de Catalunya, Barcelona, Spain\\
        E-mail: \email{jaime.haro@upc.edu}}

\author{Jaume Amor\'os
       \\ Universitat Polit\`ecnica de Catalunya, Barcelona, Spain\\
        E-mail: \email{jaume.amoros@upc.edu}  }

\abstract{It is shown that teleparallel $F(T)$  theories of 
gravity combined with holonomy corrected Loop Quantum Cosmology (LQC) support a Matter Bounce Scenario (MBS) which is a potential alternative to 
the inflationary paradigm. 
The Matter Bounce Scenario is reviewed and, according to the current observational data provided by PLANCK's team, we have summarized all the conditions that it has
to satisfy in order to be a viable alternative to inflation, such as to provide
a theoretical value of the spectral index and its running compatible with the latest PLANCK data, 
to have 
a reheating process via gravitational particle production, or to predict  some signatures in the non-gaussianities
of the power spectrum.
The calculation of the power spectrum for scalar perturbations and the ratio of tensor to scalar perturbations has been done, 
in the simplest case of an exact matter dominated background, 
for both holonomy corrected LQC and
teleparallel $F(T)$ gravity. Finally, we have discussed the challenges (essentially, dealing with non-gaussianities, the calculation of the 3-point function
in flat spatial geometries for theories beyond General Relativity) and  problems 
(Jeans instabilities in the case of holonomy corrected LQC or local Lorentz dependence in teleparallelism)
that arise in either bouncing scenario.

}

\FullConference{Frontiers of Fundamental Physics 14\\
		 15-18 July 2014\\
		 Aix Marseille University (AMU) Saint-Charles Campus, Marseille, France}

\begin{document}

\section{Introduction}
It is well-known that inflation suffers from several problems (reviewed for instance in \cite{brandenberger}), like
 the initial singularity which  is usually ignored,
   or the fine-tuning of the degree of flatness required for the potential in order to achieve successful inflation \cite{ffg91}.
   
An alternative scenario to the inflationary paradigm, called {\em Matter Bounce Scenario} (MBS), has been developed in order to
explain the evolution of our Universe \cite{cai} avoiding those problems. Essentially, it presents at very early times a matter dominated Universe in a
contracting phase, evolving towards the bounce after which it enters 
an expanding phase. This model, like inflation, solves the horizon problem that appears in General Relativity (GR) and improves the flatness problem
in GR (where spatial flatness is an unstable fixed point and  fine tuning of initial conditions is required),
because
the contribution of the spatial curvature decreases in
the contracting phase at the same rate as it increases in the expanding one (see for instance \cite{brandenberger1}).  
   
The aim of our work is to show viable bouncing cosmologies where the matter part of the Lagrangian is composed of a single scalar field and, therefore, have to 
 go beyond  General Relativity, since GR for flat spatial geometries forbids  bounces when one deals 
 with a single field. Hence,  theories such as  holonomy corrected Loop Quantum Cosmology (LQC) \cite{wilson}, where a big bounce appears owing 
 to the discrete structure of space-time \cite{ashtekar} or teleparalellism, that is $F(T)$ gravity, \cite{Haro1} must be taken into account.

The units used in the paper are: $\hbar=c=8\pi G=1$.

\section{$F({T})$ gravity in flat FLRW geometry}

Teleparallel theories are based in the {\em Weitzenb\"ock space-time}. 
This space is ${\mathcal R}^4$, with a Lorentz metric, in which a global, orthonormal basis of its tangent bundle given by 
four vector fields $\{ e_i \}$ has been selected, that is, they satisfy
$g(e_i, e_j)= \eta_{ij}$ with $\eta=\mbox{diag}\,(-1,1,1,1)$. The 
Weitzenb\"ock 
connection $\nabla$ is defined by imposing that the basis vectors $e_i$ be absolutely parallel, i.e. that $\nabla e_i=0$.

The Weitzenb\"ock connection is compatible with the metric $g$, 
and it has zero curvature because of the global parallel transport
defined by the basis $\{ e_i \}$. The information of the 
Weitzenb\"ock connection is carried by its torsion,
and its basic invariant is the {\em scalar torsion} ${T}$.
The connection, and its torsion, depend on the choice of orthonormal
basis $\{ e_i \}$, but if one adopts the flat Friedmann-Lema\^{\i}tre-Robertson-Walker (FLRW) metric and selects as orthonormal basis $\{ e_0= \partial_0,
e_1= \frac{1}{a} \partial_1, e_2= \frac{1}{a} \partial_2, 
e_3= \frac{1}{a} \partial_3 \}$,
then the scalar torsion is 
$
{T}=-6H^2$,
where $H=\frac{\dot{a}}{a}$ is the Hubble parameter,
 and this identity is invariant with respect to local Lorentz transformations that only depend on the time, i.e. of the form
$\tilde{e_i}= \Lambda^k_i(t) e_k$ (see \cite{Bengochea:2008gz,ha}).

\bigskip

With the above choice of orthonormal fields, the Lagrangian of
the $F({T})$ theory of gravity is
\begin{eqnarray}\label{eq:lagr}
{\mathcal L}_{T}={\mathcal V}(F({ T})+{\mathcal L}_M),
\end{eqnarray}
where ${\mathcal V}=a^3$ is the element of volume, 
and ${\mathcal L}_M$ is the matter Lagrangian density.

The Hamiltonian of the system is 
\begin{eqnarray}\label{eq:ham}
{\mathcal H}_{T}= \left(2T\frac{dF({T})}{dT}-F({ T})  +\rho \right){\mathcal V} \, ,
\end{eqnarray}
where ${\rho}$ is the energy density. Imposing the Hamiltonian constrain ${\mathcal H}_{ T}=0$ leads to the 
modified Friedmann equation
\begin{eqnarray}\label{eq:friedmann}
\rho=-2 \frac{dF({ T})}{dT}{T}+F({T})\equiv G({T})
\end{eqnarray}
which, as $ T=-6H^2$, defines a curve in the plane 
$(H,\rho)$.

Equation (\ref{eq:friedmann}) may be inverted, so
a curve of the form $\rho=G({T})$ defines an $F({ T})$ theory with
\begin{eqnarray}\label{eq:F(T)}
F({ T})=-\frac{\sqrt{-{ T}}}{2}\int \frac{G({ T})}{{T}\sqrt{-{ T}}}d{ T}.
\end{eqnarray}

\bigskip

To produce a cyclically evolving Universe, let us take the $F({ T})$ theory arising from the ellipse that defines the holonomy corrected Friedmann equation in Loop Quantum
Cosmology 
\begin{eqnarray}
H^2=\frac{\rho}{3}\left(1-\frac{\rho}{\rho_c}\right),
\end{eqnarray}
where $\rho_c$ is the so-called {\it critical density}.

To obtain a parametrization of the form $\rho=G({T})$,
the curve has to be split in two branches
\begin{eqnarray}\label{eq:Gpm}
\rho=G_{\pm}({ T})=\frac{\rho_c}{2}\left(1\pm\sqrt{1+\frac{2{ T}}{\rho_c}}\right),
\end{eqnarray}
where the branch $\rho=G_-({ T})$ corresponds to $\dot{H}<0$
and $\rho=G_+({ T})$ is the branch with $\dot{H}>0$.
Applying Eq. (\ref{eq:F(T)}) to these branches produces the model
(\cite{Haro1,aho13})
\begin{eqnarray}\label{eq:Fpm}
F_{\pm}({ T})=\pm\sqrt{-\frac{{ T}\rho_c}{2}}\arcsin\left(\sqrt{-\frac{2{T}}{\rho_c}}\right)+G_{\pm}({ T}).
\end{eqnarray}

\section{Matter Bounce Scenario}
Matter Bounce Scenarios (MBS) \cite{cai}  are essentially characterized by the Universe being nearly matter dominated at very early times in the contracting phase (to
obtain an approximately scale invariant power spectrum),
and  evolving 
towards a bounce where all the parts of the Universe become in causal contact \cite{aho13}, solving the horizon problem, to enter into a expanding regime, where it matches 
the behavior of the standard hot Friedmann Universe. They constitute a promising alternative to the inflationary paradigm.

According to the current observational data, in order to obtain a  viable  MBS model, the bouncing model has
to satisfy  some conditions that we have summarized  as  follows:
\begin{enumerate}
 \item
 The latest Planck data constrain the value of the spectral index for scalar perturbations and its running, namely $n_s$ and $\alpha_s$,
to $0.9603\pm 0.0073$ and $-0.0134\pm 0.009$ respectively \cite{Ade}.
The analysis of these parameters provided by Planck  makes no slow roll
approximation (for example,  the first year WMAP observations was done considering the $\Lambda$CDM model \cite{verde}), and thus,
the parameters $n_s$ and $\alpha_s$ can be used to test bouncing models.
A nearly scale invariant power spectrum of perturbations with running is obtained either as a quasi de Sitter phase in the expanding phase or as a nearly
matter domination phase at early times, in the contracting phase \cite{w99}. Then, since
for an exact matter domination in the MBS the power spectrum is flat,  to improve the model so as to match correctly with the observational data, one has to consider, at early times in the 
contracting phase,
a  quasi-matter domination period characterized by the condition $\left| w\equiv \frac{P}{\rho}\right|\ll 1$, with $P$ and $\rho$ being respectively the pressure and the energy density of the Universe.

\item The Universe has to reheat creating light particles that will thermalize matching with a hot Friedmann Universe. 
 Reheating could  be produced due to the gravitational particle creation in an
expanding Universe \cite{geometric}. In this case, an abrupt phase transition (a non adiabatic transition) is needed in order to obtain sufficient particle creation that thermalizes 
producing a reheating temperature that fits well with current observations.
It is shown in \cite{Quintin} that
gravitational particle production  could be applied to the MBS, assuming a phase transition from the matter domination
to an ekpyrotic phase, which also maintains the isotropy of the bounce, i.e., solves the Belinsky-Khalatnikov-Lifshitz instability \cite{bkl}, in the contracting regime.
In \cite{Haro2}, dealing with massless nearly conformally coupled particles, a reheating temperature compatible with current data has
recently been obtained.


\item 
The data of the seven-year survey WMAP (\cite{Larson})
constrains the value of the power spectrum for
scalar perturbations to be ${\mathcal P}_S(k)\cong 2\times 10^{-9}$.
The theoretical  results  calculated with bouncing models have to match with  that observational data.

\item
The constrain of the tensor/scalar ratio provided by WMAP and Planck projects (${r}\leq 0.11$) is obtained indirectly assuming the {\it consistency} slow roll relation
${r}=16\epsilon$ (where $\epsilon=-\frac{\dot{H}}{H^2}\cong \frac{1}{2}\left(\frac{V_{\varphi}}{V} \right)^2$ is the main slow roll parameter) \cite{Peiris}, because gravitational waves turn out not to have been
 detected by those projects. This means that the slow roll inflationary models must satisfy this constrain, but
not the bouncing ones, where there is not any consistency relation. 
Actually, to check  if the MBS models provide a viable value of the
tensor/scalar ratio, first of all gravitational waves must be clearly detected in order to determine the observed value of this ratio. 

\item 
Some non-gaussianities has been detected by PLANCK's team \cite{Adea}. A theoretical viable bouncing model has to take into account these non-gaussianities. However,
as shown by the recent calculation of the 3-point function in bouncing cosmologies \cite{Peter},
this
seems to be a very difficult challenge in MBS.

\end{enumerate}

\section{Perturbations in Matter Bounce Scenario}

The Mukhanov-Sasaki equations  for $F({ T})$ gravity and LQC are given by \cite{Cai,Cailleteau}

\begin{eqnarray}
\zeta_{S(T)}''-{c}^2_{s}\nabla^2 \zeta_{S(T)}+\frac{Z_{S(T)}'}{Z_{S(T)}}\zeta'_{S(T)}=0,
\end{eqnarray}
 where $\zeta_{S}$ and $\zeta_{T}$ denote the amplitude for scalar and tensor perturbations.
 
In  teleparallel $F({T})$ gravity one has \cite{h13}
\begin{eqnarray}
Z_S=\frac{a^2{|\Omega|}\dot{{\varphi}}^2}{{c}^2_{s}{H^2}},\quad
Z_T=\frac{a^2c^2_s}{{|\Omega|}},
\quad
c_s^2=|\Omega|\frac{
\arcsin\left(2\sqrt{\frac{3}{\rho_c}}H\right)}{2\sqrt{\frac{3}{\rho_c}}H}, \quad \mbox{with} \quad \Omega=1-\frac{2\rho}{\rho_c}.\end{eqnarray}

In contrast, for holonomy corrected  LQC one has \cite{Barrau}
\begin{eqnarray}
 Z_S=\frac{a^2\dot{{\varphi}}^2}{{H}^2},\quad
Z_T=\frac{a^2}{{\Omega}},\quad c^2_s=\Omega.
\end{eqnarray}

Dealing with the simplest model of MBS, i.e., a background depicted by an exact matter domination, whose scale factor is given by 
$a(t)=\left(\frac{3}{4}\rho_ct^2+1\right)^{1/3}$ one obtains
the following power spectrum for scalar perturbations, given by \cite{ha14}
\begin{eqnarray}
{\mathcal P}_{S}(k)=\frac{3\rho_c^2}{\rho_{pl}}
 \left|\int_{-\infty}^{\infty}{Z^{-1}_{S}(\eta)}d\eta\right|^2.
\end{eqnarray}

In the particular case of teleparalell $F({T})$  gravity one has
$ {\mathcal P}_{S}(k)
=\frac{16}{9}\frac{\rho_c}{\rho_{pl}}{\mathcal C }^2, 
$ \cite{h13}
where ${\mathcal C }=1-\frac{1}{3^2}+\frac{1}{5^2}-...=  0.915965...$ is  Catalan's constant, and for holonomy corrected LQC 
 ${\mathcal P}_{S}(k)
=\frac{\pi^2}{9}\frac{\rho_c}{\rho_{pl}} 
$ \cite{w13}.

In the same way, the ratio of tensor to scalar perturbations in MBS is given by
\begin{eqnarray}
r
=\frac{8}{3}\left(\frac{\int_{-\infty}^{\infty}{Z^{-1}_T(\eta)}d\eta}{\int_{-\infty}^{\infty}{Z^{-1}_S(\eta)}d\eta}\right)^2,
\end{eqnarray}
obtaining for teleparallel $F(T)$ gravity $r=24\left(\frac{Si(\pi/2)}{{\mathcal C}}\right)^2$, where $Si(x)\equiv\int_0^x\frac{\sin y}{y}dy$
is the Sine integral function. In contrast, for holonomy corrected LQC this ratio is very close to zero.

A final remark is in order: {\it All of the current models that depict the MBS have some problem. For example, in holonomy corrected LQC the square of the velocity of sound 
becomes negative  in the super-inflationary phase $(\rho_c/2<\rho\leq \rho_c)$. In spite of the fact that in holonomy corrected LQC, in order to calculate the power spectrum, only modes that satisfy the
long-wavelength  condition $|c_s^2k^2|\ll \left|\frac{z''}{z}\right|$ are used, it is important to realize that, in the super-inflationary regime, the ultra-violet modes 
(modes that satisfy the condition $|c_s^2k^2|\gg \left|\frac{z''}{z}\right|$)
will suffer Jeans instabilities, and thus undesirable cosmological consequences could appear. This is a serious problem that needs to be addressed in holonomy corrected LQC.
In $F(T)$  gravity, the square of the velocity of sound is always positive. However, teleparallelism suffers from the problem that the main invariant, the scalar torsion $T$, is not 
at all an invariant, in the sense that it depends on the choice of
the orthonormal basis in the tangent bundle (the tetrad).  
The scalar curvature $R$ is a real invariant, but the current bouncing models in modified $F(R)$ gravity,  do not support a matter domination in the contracting phase, and thus their power spectrum is 
not nearly scale invariant
\cite{Bamba}.

Another  possibility is to consider the MBS for a flat FLRW geometry in the context of GR. In this case one needs more than one field, one of them
a  ghost condensate field \cite{ghost} or a Galileon type field \cite{twofields} which violates the Null Energy Condition, to obtain a non-singular bounce. 
In this case the scenario is very complicated, and it is not clear at all how to compute the theoretical quantities such as the spectral index and its running, to be 
compared with the current observational data.}

\end{document}